# Selective Oxidation and Cr Segregation in High-Entropy Oxide Thin Films

Le Wang[1,†,*], Krishna Prasad Koirala[1,†], Shuhang Wu[2], Jueli Shi[1], Hsin-Mei Kao[1,3], Andrew Ho[1,4], Min-Ju Choi[1], Dongchen Qi[2], Anton Tadich[5], Mark E. Bowden[1], Bethany E. Matthews[6], Hua Zhou[7], Yang Yang[8], Chih-hung Chang[3], Zihua Zhu[9], Chongmin Wang[9], Yingge Du[1,*]

[1]Physical and Computational Sciences Directorate, Pacific Northwest National Laboratory, Richland, WA 99354, USA
[2]Centre for Materials Science, School of Chemistry and Physics, Queensland University of Technology, Brisbane, Queensland 4000, Australia
[3]School of Chemical, Biological and Environmental Engineering, Oregon State University, Corvallis, OR 97331, USA
[4]Santa Rosa Junior College, Santa Rosa, CA 95401, USA
[5]Australian Synchrotron, ANSTO, Clayton 3168, VIC, Australia
[6]Energy and Environment Directorate, Pacific Northwest National Laboratory, Richland, Washington, 99352 USA
[7]X-Ray Science Division, Advanced Photon Source, Argonne National Laboratory, Lemont, IL 60439, USA
[8]National Synchrotron Light Source II, Brookhaven National Laboratory, Upton, NY 11973, USA
[9]Environmental Molecular Sciences Laboratory, Pacific Northwest National Laboratory, Richland, WA 99354, USA

[†]These authors contributed equally to this work.

[*]Email: le.wang@pnnl.gov, yingge.du@pnnl.gov

**Abstract:** High-entropy oxides (HEOs) offer exceptional compositional flexibility and structural stability, making them promising materials for energy and catalytic applications. Here, we investigate Sr doping effects on B-site cation oxidation states, local composition, and structure in epitaxial $La_{1-x}Sr_x(Cr_{0.2}Mn_{0.2}Fe_{0.2}Co_{0.2}Ni_{0.2})O_3$ thin films. X-ray spectroscopies reveal that Sr doping preferentially promotes Cr oxidation from $Cr^{3+}$ to $Cr^{6+}$, partially oxidizes Co and Ni, while leaving $Mn^{4+}$ and $Fe^{3+}$ unchanged. Atomic-resolution scanning transmission electron microscopy with energy-dispersive X-ray spectroscopy shows pronounced Cr segregation, with Cr exhibiting depletion at the film-substrate interface and enrichment at the film surface, along with the formation of a partially amorphous phase in heavily Sr-doped samples. This segregation is likely driven by oxidation-induced migration of smaller, high-valence Cr cations during the growth. These findings underscore the critical interplay between charge transfer, local strain, and compositional fluctuations, providing strategies to control surface composition and electronic structure in HEOs for more robust electrocatalyst design.

ABO$_3$-type perovskite transition metal (*TM*) oxides are highly versatile materials due to their remarkable structural flexibility, tunable electronic properties, and wide-ranging functional applications.[1] B-site *TM* cations (such as Cr, Mn, Fe, Co, and Ni) play a crucial role in determining key material properties, such as conductivity, magnetism, and catalytic activity.[2,3] Subtle variations in B-site cation coordination, oxidation states, or *TM*-O bond symmetry can significantly influence both lattice structure and functionalities.[4-8] Building on this foundation, high-entropy perovskite oxides (HEPOs) extend the versatility of conventional perovskites by incorporating multiple equiatomic A-site and/or B-site cations, unlocking new levels of compositional complexity and functional tunability.[9-11] Recent studies on *R*(5B)O$_3$ (where R denotes a rare earth lanthanide and 5B = Cr$_{0.2}$Mn$_{0.2}$Fe$_{0.2}$Co$_{0.2}$Ni$_{0.2}$) have demonstrated promising functional characteristics, including remarkable magnetic anisotropy, enhanced exchange bias and oxygen evolution reaction (OER) activity.[12-19] These unique magnetic properties arise from competing exchange interactions among multiple cations, which promote spin frustration via energetically favorable electron configurations.[16,17] Furthermore, synergistic effects among B-site cations optimize charge transfer pathways, improving electrocatalytic performance compared to single B-site perovskites.[14] Despite these advancements, keys questions remain regarding the role of cation oxidation states, charge transfer mechanisms, and their evolution during synthesis and processing. Additionally, local composition fluctuations — a common characteristic in high-entropy oxides (HEOs) — introduce further complexity in unraveling the charge distribution mechanism in these multicomponent systems.[20-22]

Sr-doping is a well-established strategy in ABO$_3$ perovskite oxides to provides a powerful means to tune structural and electronic properties.[23-27] For example, in Sr-doped LaMnO$_3$ and LaCoO$_3$, Sr$^{2+}$ substitution for La$^{3+}$ introduces hole doping, alters B-site cation oxidation states, and enhances both magnetic and electrical properties.[23,26] Similarly, in the La$_{1-x}$Sr$_x$NiO$_{3-\delta}$ system, Sr doping shifts the O 2p band closer to the Fermi level, strengthening the Ni 3d-O 2p hybridization, reducing charge transfer energy, and significantly boosting the OER activity.[27] However, Sr-doping can also lower the oxygen vacancy formation energy in La*TM*O$_3$ systems,[28-30] potentially leading to secondary phase formation and cation leaching during electrochemical reactions,[31-33] which may undermine materials stability. While these effects are well studied in conventional perovskites with single or dual B-site cations, their impact on HEPOs like La(5B)O$_3$ remains largely unexplored.[34-36] Critical unresolved questions include: How does Sr-doping influence oxidation states and charge transfer among multiple B-site cations (Figure 1a)? Does Sr doping induce oxygen vacancies in these multicomponent systems? And how do local composition fluctuations and potential secondary phases collectively affect stability and performance?

In this work, we investigate the effects of Sr-doping on the charge distribution and composition fluctuations in epitaxial La(5B)O$_3$ thin films synthesized by pulsed laser deposition (PLD). Leveraging advanced spectroscopy techniques, including X-ray photoelectron spectroscopy (XPS), soft X-ray absorption spectroscopy (XAS), and hard XAS, we elucidate cation-specific oxidation state changes induced by Sr incorporation. Among the five B-site *TM* elements, Mn is the most prone to oxidation, stabilizing in a Mn$^{4+}$ state (d$^3$ configuration) even in the undoped state. With Sr doping, Cr preferentially oxidizes from predominantly Cr$^{3+}$ to Cr$^{6+}$, followed by partial oxidation of Co and Ni, while Fe remains in a stable Fe$^{3+}$ state. At 50% Sr doping, Co undergoes a spin state transition from low-spin to high-spin, further modulating the electronic structure. Additionally, scanning transmission electron microscopy (STEM) combined with atomic-resolution energy-dispersive X-ray spectroscopy (EDS) reveals the spatially non-uniform distribution of cations, highlighting Cr segregation with Cr exhibiting depletion near the film/substrate interface while becoming enriched at the film surface, further confirmed by time-of-flight secondary ion mass spectrometry (ToF-SIMS) measurements. Layer-resolved electron energy-loss spectroscopy (EELS) confirms that Cr is present as Cr$^{3+}$ at the interface while the oxidation state increases along the growth direction. It is also found that increasing Sr doping intensifies Cr segregation. These observations of Cr segregation and cation redistribution provide deeper insights into the complexity of HEOs, refining our understanding of how local composition fluctuations influence structural, electronic, and potentially catalytic properties.

Epitaxial La$_{1-x}$Sr$_x$(5B)O$_3$ (where 5B = Cr$_{0.2}$Mn$_{0.2}$Fe$_{0.2}$Co$_{0.2}$Ni$_{0.2}$) thin films with x = 0, 0.25, and 0.5 and thicknesses ranging from 10 to 25 nm were grown on (001)-oriented (LaAlO$_3$)$_{0.3}$(Sr$_2$AlTaO$_6$)$_{0.7}$ (LSAT) substrates using PLD (see Methods for details). X-ray diffraction (XRD) θ−2θ scans (Figure 1b) show distinct Laue fringes for x = 0 and x = 0.25, indicating high crystalline quality. For x = 0.5, the fringes are less pronounced, suggesting increased structural disorder. Reciprocal space maps (RSMs) (Figure 1c) confirm that all films remained coherently strained to the substrate. Notably, the (001) diffraction peak systematically shifts to higher diffraction angles with increasing Sr content, suggesting a decrease in the out-of-plane lattice parameter, consistent with previous reports.[34] This trend aligns with the expected effect of Sr substitution in La$TM$O$_3$ thin films,[23,24] where replacing the larger La$^{3+}$ ions with smaller Sr$^{2+}$ ions reduces the average A-site ionic radii, causing lattice contraction and shorter $TM$-O bond lengths. Additionally, the oxidation of B-site cations induced by Sr doping further contributes to the lattice contraction,[24,25] as higher-valence cations generally have smaller ionic radii.

To examine how Sr substitution influences the oxidation states of B-site cations in La$_{1-x}$Sr$_x$(5B)O$_3$, XPS and XAS measurements were performed. Figure 2a shows the $TM$ 3p core level XPS spectra for x = 0, 0.25, and 0.5. At x = 0, distinct features corresponding to all $TM$s are clearly observed, but increasing Sr content systematically weakens the Cr 3p signal and intensifies the Mn 3p signal. This trend indicates an increase in the oxidation state of Cr from Cr$^{3+}$ to Cr$^{6+}$ (corroborated by Cr 2p XPS spectra; Figure S1), shifting its 3p binding energy closer to that of Mn and increasing the overall apparent area of the Mn 3p component. Meanwhile, Ni and Co 3p peaks shift slightly to higher binding energies, suggesting their oxidation (supported by Co and Ni $K$-edge XAS spectra; Figure S2) with Sr doping. In contrast, the Mn 3p and Fe 3p binding energies remain unchanged, implying stable Mn and Fe oxidation states, as confirmed by the Mn 2p XPS spectra (Figure S1) and Fe $K$-edge XAS spectra (Figure S2). These selective oxidation behaviors align with earlier findings by Kante et al.,[14] who observed similar trends via ambient pressure XPS in La(5B)O$_3$ thin films.

To further elucidate the impact of Sr doping on B-site cation oxidation states in La$_{1-x}$Sr$_x$(5B)O$_3$, we performed $TM$ $L$-edge XAS measurements. Figure 2b presents the Cr $L$-edge, Mn $L$-edge, Fe $L_3$-edge, Co $L_3$-edge, and Ni $L_2$-edge spectra for La$_{1-x}$Sr$_x$(5B)O$_3$ thin films, along with reference spectra for comparison and quantitative analysis. For x = 0, Cr shows a mixed Cr$^{3+}$/Cr$^{4+}$/Cr$^{6+}$ state that shifts increasingly toward Cr$^{6+}$ with higher Sr content. Meanwhile, Mn remains as Mn$^{4+}$ at all doping levels. This stability is likely due to the strong crystal field stabilization energy of Mn$^{4+}$ (d$^3$, $t_{2g}^3$ configuration) compared to Mn$^{3+}$ (d$^4$, $t_{2g}^3 e_g^1$ configuration). Similar behavior has been observed in other systems, such as the double perovskites La$_2$MnNiO$_3$ and La$_2$MnCoO$_3$.[37-39]

The Fe $L_3$-edge XAS spectra exhibit notable changes with Sr doping. For x = 0, the Fe $L_3$-edge feature shows a well-resolved double-peak structure, which becomes less defined as Sr content increases. While these changes could suggest variations in the oxidation state, previous studies have showed that the Fe $L$-edge features are also strongly influenced by the local structure and coordination environment. Instead of relying solely on peak shapes, comparisons of peak widths provide a more reliable method for determining the Fe oxidation state.[40,41] By comparing the Fe $L_3$ spectra of the x = 0.5 sample with that of $SrFe^{3+}O_{2.5}$,[42] a strong spectral overlay confirms that Fe remains in the $Fe^{3+}$ oxidation state for x = 0.5. This conclusion is consistent with Fe 3p XPS (Figure 2a) and Fe $K$-edge XAS results (Figure S2). The stability of $Fe^{3+}$ is attributed to its half-filled 3d orbital ($d^5$, $t_{2g}^3 e_g^2$ configuration), which is one of the most stable electronic arrangements in *TM* cations.

The Co $L_3$-edge spectra exhibit a mixture of $Co^{2+}$ and $Co^{3+}$ for all $La_{1-x}Sr_x(5B)O_3$ films. For x = 0, the spectrum primarily consists of contributions from $Co^{2+}$ and low spin (LS) $Co^{3+}$. With Sr doping (x = 0.25), the LS $Co^{3+}$ feature becomes more pronounced, indicating a shift toward higher oxidation states. At x = 0.5, a spin-state transition to high-spin (HS) $Co^{3+}$ is observed, which could significantly alter the magnetic properties of the films.[34] The average Co oxidation state, determined by fitting the experimental spectra to a linear combination of $Co^{2+}$ and $Co^{3+}$ reference spectra, is summarized in Figure 2c.

For Ni, the x = 0 sample closely matches the $Ni^{2+}$ reference. With increasing Sr doping level, the intensity of peak α decreases while peak β becomes more pronounced, suggesting a progressive increase in the Ni oxidation state.[41,43-45] Using the peak area ratio $I_α/(I_α + I_β)$ and a reference value of 0.62 for $Ni^{2+}$, the Ni oxidation states were calculated and are shown in Figure 2c, confirming a shift toward higher oxidation states with increasing Sr content. This trend aligns with Ni 3p XPS (Figure 2a) and Ni $K$-edge XAS results (Figure S2).

Figure 2c summarizes these quantitative analyses. For x = 0, the average oxidation states are $Cr^{3.4+}$, $Mn^{4+}$, $Fe^{3+}$, $Co^{2.6+}$, and $Ni^{2+}$. These values maintain overall charge neutrality in the $La(5B)O_3$ lattice but deviate slightly from previous studies, such as those by Wang *et al.*[19] ($Cr^{3+}$, $Mn^{4+}$, $Fe^{3+}$, $Co^{3+}$, and $Ni^{2+}$) for thin films grown on $SrTiO_3$ substrates and Nguyen *et al.*[46] ($Cr^{3+/6+}$, $Mn^{4+/3+}$, $Fe^{3+/2+}$, $Co^{3+/2+}$, and $Ni^{2+}$) for bulk samples. These differences can be attributed to a combination of factors, including dimensionality effects (thin film vs. bulk), synthesis conditions (such as deposition temperature and oxygen partial pressure), substrate effects, and the inherent chemical and structural flexibility of HEPOs. With Sr doping (x = 0.25 or 0.5), significant changes are observed: Cr oxidizes substantially to $Cr^{6+}$, and both Co and Ni show increased oxidation states. In contrast, Mn and Fe remain stable as $Mn^{4+}$ and $Fe^{3+}$, respectively. Such charge distribution cannot be explained by electronegativity alone (as a purely electronegativity-based argument would predict oxidation in the order Cr > Mn > Fe > Co > Ni with Sr doping), but rather results from a complex interplay of factors: (1) crystallinity and structural stability, which favor Jahn-Teller-inactive ions (e.g., $Mn^{4+}$, $Fe^{3+}$) in high-symmetry lattices; (2) charge balance, where $Sr^{2+}$ doping drives redox flexibility in Cr, Co, and Ni; and (3) local strain effects, whereby oxidation-state changes alter cation radii and *TM*–O bond lengths, propagating further adjustments in neighboring coordination environments. This dynamic interplay underlies the rich oxidation-state behavior of high-entropy perovskite oxides.

To provide a more detailed characterization of the crystal structure and cation distribution in these HEPO thin films, we performed STEM and EDS measurements and analyses. As shown in Figure 3a, an atomically resolved view of a 20 nm-thick La(5B)$O_3$ (x = 0) film grown on LSAT reveals its high crystalline quality and a sharp interface. EDS maps (Figure S3) and integrated line profiles (Figure 3b) illustrate the spatial distribution of elements, highlighting Cr segregation. Enlarged HAADF-STEM images in Figures 3c and 3d (corresponding to regions marked by green and yellow boxes in Figure 3a, respectively) and their accompanying EDS maps clearly demonstrate Cr depletion at the film/substrate interface and Cr enrichment at the film surface (Figure 3b). To further examine how this segregation affects cation oxidation states, we conducted layer-resolved EELS measurements (Figure S4). Among all the cations studied, only Cr exhibits a progressive increase in oxidation state, transitioning from predominantly $Cr^{3+}$ at the interface to a mixed $Cr^{3+/4+}$ state near the surface, as evidenced by shifts in Cr $L_3$-edge peak positions. These observations suggest that the Cr segregation is closely linked to its redox-mediated oxidation during PLD growth. Recent studies on HEOs have shown that the oxidation state and coordination preference of Cr are highly sensitive to oxygen partial pressure.[47] Under the high oxygen partial pressure (150 mTorr) utilized to synthesize La$_{1-x}$Sr$_x$(5B)$O_3$ films, Cr readily oxidizes to higher-valence states beyond $Cr^{3+}$. For x = 0, we determined the average Cr valence to be $Cr^{3.4+}$, using a linear combination of $Cr^{3+}$, $Cr^{4+}$ and $Cr^{6+}$. The formation of $Cr^{4+}$ and $Cr^{6+}$, which have smaller ionic radii than $Cr^{3+}$, shortens Cr–O bond lengths, inducing localized lattice strain. As this strain propagates, it accumulates elastic energy within the lattice, increasing the system's free energy. To relieve elastic energy and minimize free energy, the lattice undergoes structural relaxation, redistributing stress and expelling the smaller high valence Cr cations to energetically favorable regions (such as the film surface), ultimately driving Cr segregation. Sr doping will amplify this effect, as $Sr^{2+}$ substitution for $La^{3+}$ promotes further Cr oxidation (Figure 2c), increasing ionic radii disparities and strain.

As illustrated in Figure 4, Sr doping in La(5B)O$_3$ significantly enhances Cr segregation. Compared to the relatively flat surface of the undoped La(5B)O$_3$ film (Figure 3), the La$_{1-x}$Sr$_x$(5B)O$_3$ (x = 0.5) sample exhibits a noticeably rougher surface (Figure 4a), along with amorphous regions (Figure 4b). EDS maps (middle and right panels of Figure 4b) and integrated line profiles (Figure 4c) reveal a reduced concentration of Cr at the film–substrate interface, with Cr enrichment in the amorphous regions. These regions also exhibit Sr enrichment, suggesting the formation of SrCrO$_4$ with Cr in the Cr$^{6+}$ oxidation state. An enlarged HAADF-STEM image (light blue box in Figure 4b) in Figure 4d, along with corresponding EDS maps, further confirms pronounced Cr depletion at the film–substrate interface. In regions devoid of amorphous structures (blue box in Figure 4a), additional EDS maps (Figure S5) also indicate Cr depletion at the interface but no detectable Cr enrichment at the film surface. These findings indicate that Cr migration occurs, in tandem with Sr segregation, leading to the formation of island-like amorphous aggregates. Furthermore, ToF-SIMS line profiles (Figure S6) confirm progressive Cr segregation, showing the broadening of the Cr depletion layers as the Sr doping level increases from 0% to 50%.

The formation of Cr$^{6+}$ in Sr-doped samples raises important concerns regarding phase instability and material degradation. Cr$^{6+}$-containing phases, such as SrCrO$_4$, are often thermodynamically metastable in reducing or humid environments, making them susceptible to decomposition or dissolution, which can accelerate structural deterioration. Additionally, Cr$^{6+}$ is known to be highly toxic and environmentally hazardous,[48-50] raising potential concerns for applications where material leaching or exposure to moisture could lead to contamination. These findings highlight the critical need for precise control over oxidation conditions and doping strategies to prevent undesired phase segregation and stability issues.

In summary, we elucidate the impact of Sr doping on cation-specific oxidation states and segregation distribution in epitaxial La$_{1-x}$Sr$_x$(5B)O$_3$ (where 5B = Cr$_{0.2}$Mn$_{0.2}$Fe$_{0.2}$Co$_{0.2}$Ni$_{0.2}$) thin films. Using a combination of XPS and XAS spectroscopies, we demonstrate selective B-site oxidation: Cr undergoes preferential oxidation from Cr$^{3+}$ to Cr$^{6+}$, while Mn (Mn$^{4+}$) and Fe (Fe$^{3+}$) remain stable due to their rigid d$^3$ and d$^5$ electronic configurations, respectively. Co and Ni exhibit increased oxidation states, with Co undergoing a low-spin to high-spin transition at 50% Sr doping. These trends deviate from predictions based solely on electronegativity, reflecting a complex interplay of crystallinity and structural stability, charge balance, and local strain effects. Importantly, Cr segregation is observed in these B-site HEO thin films, intensifying with Sr content. STEM/EDS and ToF-SIMS reveal Cr depletion at the film-substrate interface and enrichment at the surface, forming $SrCrO_4$-rich amorphous regions for Sr-doped samples. The observed Cr segregation is likely driven by redox-induced ionic radii mismatch and strain relief during the growth. Future work will focus on the role of additional synthesis parameters (e.g., oxygen partial pressure and laser fluence) in controlling cation segregation. Our findings emphasize that a deeper understanding of metastable phase formation and segregation mechanisms is crucial for establishing composition–structure–stability–property relationships in HEOs, ultimately unlocking advanced functional opportunities.

## ASSOCIATED CONTENT

### Supporting Information

This supporting information is available free of charge via the internet at http://pubs.acs.org.

Methods; Cr 2p and Mn 2p XPS data; Fe K-edge, Co K-edge, and Ni K-edge XAS data; HAADF-STEM images and EDS elemental maps of the La(5B)$O_3$/LSAT interface; layer-resolved EELS of the O K-edge, Cr L-edge, Mn L-edge, Fe L-edge, and Co L-edge spectra; HAADF-STEM images and EDS elemental maps of the $La_{0.5}Sr_{0.5}$(5B)$O_3$/LSAT interface; and ToF-SIMS results.

## AUTHOR INFORMATION

### Corresponding Authors
*Email: le.wang@pnnl.gov; yingge.du@pnnl.gov### Notes

†These authors contributed equally. The authors declare no competing financial interest.

## ACKNOWLEDGEMENTS

This work was supported by the U.S. Department of Energy (DOE), Office of Science, Basic Energy Sciences, Division of Materials Sciences and Engineering, Synthesis and Processing Science Program, FWP 10122. We sincerely thank Dr. Peter V. Sushko and Dr. Krishna Chaitanya Pitike for their valuable discussions and insights in interpreting the experimental results. Soft XAS measurements were performed at the Soft X-ray beamline (SR14ID01) at the Australian Synchrotron. Hard XAS measurements used the Submicron Resolution X-ray Spectroscopy (SRX) beamline at 5-ID of the National Synchrotron Light Source II, a U.S. DOE Office of Science User Facility operated for the DOE Office of Science by Brookhaven National Laboratory under Contract No. DE-SC0012704. Part of the STEM and focused ion beam SEM work was carried out using microscopes that are funded by a grant from the Washington State Department of Commerce's Clean Energy Fund. A portion of work was performed via a project award (Award DOI: 10.46936/cpcy.proj.2021.60271/60008423) from the Environmental Molecular Sciences Laboratory, a DOE Office of Science User Facility sponsored by the Biological and Environmental Research program under Contract No. DE-AC05-76RL01830.

**References**

1       Peña, M. A.; Fierro, J. L. Chemical Structures and Performance of Perovskite Oxides. *Chem. Rev.* **2001**, *101*(7), 1981-2018.
2       Suntivich, J.; May, K. J.; Gasteiger, H. A.; Goodenough, J. B.; Shao-Horn, Y.; A Perovskite Oxide Optimized for Oxygen Evolution Catalysis from Molecular Orbital Principles. *Science* **2011**, *334*(6061), 1383-1385.
3       Imada, M.; Fujimori, A.; Tokura, Y. Metal-insulator Transitions. *Rev. Mod. Phys.* **1998**, *70*(4), 1039.
4       Huang, T.; Lyu, Y.; Huyan, H.; Ni, J.; Saremi, S.; Wang, Y.; Yi, D.; He, Q.; Martin, L.W.; Xiang, H. Manipulation of the Ferromagnetism in $LaCoO_3$ Thin Films Through Cation-Stoichiometric Engineering. *Adv. Electron. Mater.* **2023**, *9*(5), 2201245.
5       Xu, Z.T.; Jin, K.J.; Gu, L.; Jin, Y.L.; Ge, C.; Wang, C.; Guo, H.Z.; Lu, H.B.; Zhao, R.Q.; Yang, G.z. Evidence for a Crucial Role Played by Oxygen Vacancies in $LaMnO_3$ Resistive Switching Memories. *Small* **2012**, *8*(8), 1279-1284.
6       An, Q.; Xu, Z.; Wang, Z.; Meng, M.; Guan, M.; Meng, S.; Zhu, X.; Guo, H.; Yang, F.; Guo, J. Tuning of the Oxygen Vacancies in $LaCoO_3$ Films at the Atomic Scale. *Appl. Phys. Lett.* **2021**, *118*, 081602.
7       Wang, L.; Dash, S.; Chang, L.; You, L.; Feng, Y.; He, X.; Jin, K.J.; Zhou, Y.; Ong, H.G. Ren, P.; Wang, S. Oxygen Vacancy Induced Room-temperature Metal–Insulator Transition in Nickelate Films and Its Potential Application in Photovoltaics. *ACS Appl. Mater. Interfaces.* **2016**, *8*(15), 9769-9776.
8       Kotiuga, M. Zhang, Z.; Li, J. ; Rodolakis, F.; Zhou, H.; Sutarto, R.; He, F.; Wang, Q.; Sun, Y.; Wang, Y.; Aghamiri, N.A. Carrier Localization in Perovskite Nickelates from Oxygen Vacancies. *Proc. Natl. Acad. Sci.* **2019**, 116(44), 21992-21997.
9       Ma, J.; Liu, T.; Ye, W.; He, Q.; Chen, K.; High-entropy Perovskite Oxides for Energy Materials: A Review. *J. Energy Storage* **2024**, *90*, 111890.

10  Wang, L.; Hossain, M. D.; Du, Y.; Chambers, S. A. Exploring the Potential of High Entropy Perovskite Oxides as Catalysts for Water Oxidation. *Nano Today* **2022**, *47*, 101697.

11  Jiang, S.; Hu, T.; Glid, J.; Zhou, N.; Nie, J.; Qin, M.; Harrinton, T.; Vecchio, K.; Luo, J.; A New Class of High-Entropy Perovskite Oxides. *Scr. Mater.* **2018**, *142*, 116-120.

12  Regmi, B. Cocconcelli, M.; Miertschin, D.; Salinas, D.P.; Panchal, G.; Kandel, P.; Pandey, K.; Ogunniranye, I.; Mueller, R.; Yao, L.; Valvidares, M.; Epitaxial Growth and Magnetic Characterization of Orthorhombic Ho($Ni_{0.2}Co_{0.2}Fe_{0.2}Mn_{0.2}Cr_{0.2}$)$O_3$ High-Entropy Oxide Perovskite Thin Films. *J. Magn. Magn. Maer.* **2025**, *613*, 172673.

13  Li, W.; Zhao, Z.; Zhao, J.; Wang, Y.; Wang, X. High Entropy La($Cr_{0.2}Mn_{0.2}Fe_{0.2}Co_{0.2}Ni_{0.2}$)$O_3$ with Tailored $e_g$ Occupancy and Transition Metal-Oxygen Bond Properties for Oxygen Reduction Reaction. *J. Mater. Sci. Technol.* **2024**, *194*, 236-246.

14  Kante, M. V.; Wever, M.L.; Ni, S.; van denBisch, I.C.; van der Minne, E.; Heymann, L.; Falling, L.J.; Gauquelin, n.; Tsvetanova, M.; Cunha, M.; Koster, G.A. A High-Entropy Oxide as High-Activity Electrocatalyst for Water Oxidation. *ACS nano* **2023**, *17*(6), 5329-5339.

15  Farhan, A.; Cocconcelli, M.; Stramaglia, F.; Kuznetsov, N.; Flajsman, L.; Wyss, M.; Yao, L.; Piamonteze, C.; and Van Dijken, S. Element-Sensitive X-ray Absorption Spectroscopy and Magnetometry of Lu($Fe_{0.2}Mn_{0.2}Co_{0.2}Cr_{0.2}Ni_{0.2}$)$O_3$ High-Entropy Oxide Perovskite Thin Films. *Phys. Rev. Mater.* **2023**, *7*(4), 044402.

16  Farhan, A.; Stramaglia, F.; Cocconcelli, M.; Kuznetsov, N.; Yao, L.; Kleibert, A.; Piamonteze, C.; Van Dijken, S.Weak ferromagnetism in Tb($Fe_{0.2}Mn_{0.2}Co_{0.2}Cr_{0.2}Ni_{0.2}$)$O_3$ High-Entropy Oxide Perovskite Thin Films. *Phys. Rev. B* **2022**, *106*(6), L060404.

17  Sharma, Y.; Zheng, Q.; Mazza, A.R.; Skoropata, E.; Heitmann, T.; Gai, Z.; Musico, B.; Miceli, P.F.; Sales, B.C.; Keppens, V.; Brahlek, M. Magnetic Anisotropy in Single-Crystal High-Entropy Perovskite Oxide La($Cr_{0.2}Mn_{0.2}Fe_{0.2}Co_{0.2}Ni_{0.2}$)$O_3$ Films. *Phys. Rev. Mater.* **2020**, *4*(1), 014404.

18  Brahlek, M.; Mazza, A.R.; Pitike, K.C.; Skoropata, E.; Lapano, J.; Eres, G.; Cooper, V.R.; Ward, T.Z. Unexpected Crystalline Homogeneity from the Disordered Bond Network in La($Cr_{0.2}Mn_{0.2}Fe_{0.2}Co_{0.2}Ni_{0.2}$)$O_3$ Films. *Phys. Rev. Mater*. **2020**, *4*(5), 054407.

19  Wang, H.; Huang, H.; Feng, Y.; Ku, Y.C.; Liu, C.E.; Chen, S.; Farhan, A.; Piamonteze, C.; Lu, Y.; Tang, Y.; Wei, J. Enhanced Exchange Bias in Epitaxial High-Entropy Oxide Heterostructures. *ACS Appl. Mater. Interfaces*. **2023**, *15*(50), 58643-58650.

20  Su, L.; Huyan, H.; Sarkar, A.; Gao, W.; Yan, X.; Addiego, C.; Kruk, R.; Hahn, H.; Pan, X. Direct Observation of Elemental Fluctuation and Oxygen Octahedral Distortion-Dependent Charge Distribution in High Entropy Oxides. *Nat. Commun.* **2022**, *13*(1), 2358.

21  Gao, H.; Guo, N.; Gong, Y.; Bai, L.; Wang, D.; Zheng, Q. Sub-Ångstrom-Scale Structural Variations in High-Entropy Oxides. *Nanoscale* **2023**, *15(*48), 19469-19474.

22  Patel, R. K.; Jenjeti, R.N.; Kumar, R.; Bhattacharya, N.; Kumar, S.; Ojha, S.K.; Zhang, Z.; Zhou, H.; Qu, K.; Wang, Z.; Yang, Z. Thickness Dependent OER Electrocatalysis of Epitaxial Thin Film of High Entropy Oxide. *Appl. Phys. Rev.* **2023**, *10(*3), 031407.

23  Shen, Z.; Qu, M.; Shi, J.; Oropeza, F.E.; Gorni, G.; Tian, C.M.; Hofmann, J.P.; Cheng, J.; Li, J.; Zhang, K.H. Correlating the Electronic Structure of Perovskite $La_{1-x}Sr_xCoO_3$ with Activity for the Oxygen Evolution Reaction: The Critical Role of Co 3d Hole State. *J. Energy Chem.* **2022**, *65*, 637-645.

24  Wang, L.; Du, Y.; Sushko, P.V. ; Bowden, M.E.; Stoerzinger, K.A.; Heald, S.M.; Scafetta, M.D.; Kasper, T.C.; Chambers, S.A. Hole-Induced Electronic and Optical Transitions in $La_{1-x}Sr_xFeO_3$ Epitaxial Thin Films. *Phys. Rev. Mater.* **2019**, *3*(2), 025401.

25  Zhang, K. H.; Du, Y.; Sushko, P.; Bowden, M.E.; Shutthandandan, V. ; Sallis, S.; Piper, L.F.; Chambers, S.A. Hole-Induced Insulator-to-Metal Transition in $La_{1-x}Sr_xCrO_3$ Epitaxial Films. *Phys. Rev. B* **2015**, *91*(15), 155129.

26  Chaluvadi, S. K.; Polewczyk, V. ; Petrov, A.Y.; Vinai, G.; Braglia, L.; Diez, J.M.; Pierron, V.; Perna, P.; Mechin, L.; Torelli, P.; Orgiani, P. Electronic Properties of Fully Strained $La_{1-x}Sr_x MnO_3$ Thin Films Grown by Molecular Beam Epitaxy ($0.15 \leq x \leq 0.45$). *ACS omega* **2022**, 7(17), 14571-14578.

27  Liu, J.; Jia, E.; Wang, L.; Stoerzinger, K.A.; Zhou, H.; Tang, C.S.; Yin, X.; Bousquet, E.; Bowden, M.E.; Wee, A.T. Tuning the Electronic Structure of LaNiO$_3$ Through Alloying With Strontium to Enhance Oxygen Evolution Activity. *Adv. Sci.* **2019**, *6*(19), 1901073.

28  Wexler, R. B.; Gautam, G. S.; Stechel, E. B.; Carter, E. A. Factors Governing Oxygen Vacancy Formation in Oxide Perovskites. *J. Am. Chem. Soc.* **2021**, *143*(33), 13212-13227.

29  Sun, M.; He, X.; Chen, M.; Tang, C.S.; Liu, X.; Dai, L.; Liu, J.; Zeng, Z.; Sun, S.; Breese, M.B.; Cai, C. Tunable Collective Excitations in Epitaxial Perovskite Nickelates. *ACS Photonics* **2024**, *11*(6) 2324-2334.

30  Feng, Y.; Jin, H.; Wang, S. Oxygen Migration Performance of LaFeO$_3$ Perovskite-type Oxygen Carriers with Sr Doping. *Phys. Chem. Chem. Phys.* **2023,** *25*(13), 9216-9224.

31  Sankannavar, R.; Sandeep, K.; Kamath, S.; Suresh, A. K.; Sarkar, A. Impact of Strontium-Substitution on Oxygen Evolution Reaction of Lanthanum Nickelates in Alkaline Solution. *J. Electrochem. Soc.* **2018**, 165(15), J3236.

32  Mefford, J. T.; Rong, X.; Abakumov, A.M.; Hardin, W.G.; Dai, S.; Kolpak, A.M.; Johnston, K.P.; Stevenson K. J. Water Electrolysis on La$_{1-x}$Sr$_x$CoO$_{3-\delta}$ Perovskite Electrocatalysts. *Nat. commun.* **2016,** *7*(1), 11053.

33  Adiga, P.; Wang, L.; Wong, C.; Matthews, B.E.; Bowden, M.E.; Spurgeon, S.R.; Sterbinsky, G.E. Blum, M.; Choi, M.J.; Tao, J.; Kasper, T.C.; Chambers, S.A.; Stoerzinger, K.A.; Du, Y. Correlation Between Oxygen Evolution Reaction Activity and Surface Compositional Evolution in Epitaxial La$_{0.5}$Sr$_{0.5}$Ni$_{1-x}$Fe$_x$O$_{3-\delta}$ Thin Films. *Nanoscale* **2023**, *15*(3) 1119-1127.

34  Mazza, A. R.; Skoropata, E.; Lapano, J.; Zhang, J.; Sharma, Y.; Musico, B.L.; Keppens, V.; Gai, Z.; Brahlek, M.J.; Moreo, A.; Gilbert, D.A.; Dagotto, E.; Ward, T.Z. Charge Doping Effects on Magnetic Properties of Single-Crystal La$_{1-x}$Sr$_x$(Cr$_{0.2}$Mn$_{0.2}$Fe$_{0.2}$Co$_{0.2}$Ni$_{0.2}$)O$_3$ (0≤x≤0.5) High-Entropy Perovskite Oxides. *Phys. Rev. B,* **2021**, *104*(9), 094204.

35  Li, M.; Zhi, Q.; Li, J.; Wu, C.; Jiang, X.; Zhang, R.; Wang, H.; Wang, H.; Fan, B. Sr(Cr$_{0.2}$Mn$_{0.2}$Fe$_{0.2}$Co$_{0.2}$Ni$_{0.2}$)O$_3$: A Novel High-Entropy Perovskite Oxide with Enhanced Electromagnetic Wave Absorption Properties. *J. Materiomics,* **2024**, 10(6), 1176-1185.

36  Liu, Z.; Xu, H.; Wang, X. ; Tian, G.; Du, D. ; Shu, C. Strain-Rich High-Entropy Perovskite Oxide of (La$_{0.8}$Sr$_{0.2}$)(Mn$_{0.2}$Fe$_{0.2}$Cr$_{0.2}$Co$_{0.2}$Ni$_{0.2}$)O$_3$ for Durable and Effective Catalysis of Oxygen Redox Reactions in Lithium-Oxygen Battery. *Battery Energy* **2024,** *3*(2), 20230053.

37  Gauvin-Ndiaye, C.; Baker, T.E.; Karan, P.; Masse, E.; Balli, M.; Brahiti, N.; Eskandari, M.A.; Fourier, P.; Tremblay, A.M.; Nourafkan, R. Electronic and Magnetic Properties of the Candidate Magnetocaloric-Material Double Perovskites La$_2$MnCoO$_6$, La$_2$MnNiO$_6$, and La$_2$MnFeO$_6$. *Phys. Rev. B* **2018**, *98*(12), 125132.

38  Guo, H.; Gupta, A.; Varela, M.; Pennycook, S.; Zhang, J. Local Valence and Magnetic Characteristics of La$_2$NiMnO$_6$. *Phys. Rev. B Condens. Matter.* **2009**, *79*(17), 172402.

39  Burnus, T.; Hu, Z.; Hsieh, H.H.; Joly, V.J.; Joy, P.A.; Haverkort, M.W.; Wu, H.; Tanaka, A.; Lin, H.J.; Chen, C.T.; Tjeng, L.H. Local Electronic Structure and Magnetic Properties of LaMn$_{0.5}$Co$_{0.5}$O$_3$ Studied by X-ray Absorption and Magnetic Circular Dichroism Spectroscopy. *Phys. Rev. B Condens. Matter.* **2008**, *77*(12), 125124.

40  Wang, L.; Yang, Z.; Koirala, K.P.; Bowden, M.E.; Freeland, J.W.; Sushko, P.V.; Kuo, C.T.; Chambers, S.A.; Wang, C.; Jalan, B.; Du, Y. Interfacial Charge Transfer and its Impact on Transport Properties of LaNiO$_3$/LaFeO$_3$ Superlattices. *Sci. Adv.* **2024,** *10*(51), eadq6687.

41  Wang, L.; Adiga, P.; Zhao, J.; Samarakoon, W.S.; Stoerzinger, K.A.; Spurgeon, S.R.; Matthews, B.E.; Bowden, M.E.; Sushko, P.V.; Kaspar, T.C.; Sterbinsky, G.E. Understanding the Electronic Structure Evolution of Epitaxial LaNi$_{1-x}$Fe$_x$O$_3$ Thin Films for Water Oxidation. *Nano Lett.* **2021**, *21*(19), 8324-8331.

42  Wang, L. Yang, Z.; Wu, J.; Bowdend, M.E.; Yang, W.; Qiao, A.; Du, Y. Time-and Strain-Dependent Nanoscale Structural Degradation in Phase Change Epitaxial Strontium Ferrite Films. *npj Mater. Degrad.* **2020,** *4*(1), 16.

43  Liu, X.; Kotiuga, M.; Kim, H.S.; N'Diaye, A.T.; Choi, Y.; Zhang, Q.; Cao, Y.; Kareev, M.; Wen, F.; Pal, B.; Freeland, J.W. Interfacial Charge-Transfer Mott State in Iridate–Nickelate Superlattices. *Proc. Natl. Acad. Sci.* **2019**, *116*(40), 19863-19868.

44  Wang, L.; Stoerzinger, K.A.; Chang, L.; Zhao, J.; Li, Y.; Tang, C.S.; Yin, X.; Bowden, M.E.; Yang, Z.; Guo, H.; You, L.; Guo, R.; Wang, J.; Ibrahim, K.; Chen, J.; Rusydi, A.; Wang, J.; Chambers, S.A.; Du, Y. Tuning Bifunctional Oxygen Electrocatalysts by Changing the A-Site Rare-Earth Element in Perovskite Nickelates. *Adv. Funct. Mater.* **2018**, *28*(39), 1803712.

45  Guo, E. J.; Liu, Y.; Sohn, C.; Desautels, R.D.; Herklotz, A.; Liao, Z.; Nichols, J.; Freeland, J.W.; Fitzsimmons, M.R.; Lee, H.N.; Oxygen Diode Formed in Nickelate Heterostructures by Chemical Potential Mismatch. *Adv. Mater.* **2018**, *30*(15), 1705904.

46  Nguyen, T. X.; Liao, Y. C.; Lin, C. C.; Su, Y. H.; Ting, J. M. Advanced High Entropy Perovskite Oxide Electrocatalyst for Oxygen Evolution Reaction. *Adv. Funct. Mater.* **2021**, *31*(27), 2101632.

47  Niculescu, G. E.; Bejger G.R.; Barber, J.P.; Wright, J.T.; Almishal, S.S.; Webb, M.; Ayyagari, S.V.G.; Maria, J.P.; Alem, N.; Heron, J.T.; Rost, C.M. Local Structure Maturation in High Entropy Oxide $(Mg,Co,Ni,Cu,Zn)_{1-x}(Cr,Mn)_xO$ Thin Films. *J Am. Ceram. Soc.* **2024**, *108*(2), e20171.

48  Yan, G.; Gao, Y.; Xue, K.; Qi, Y.; Fan, Y.; Tian, X.; Wang, J.; Zhao, R.; Zhang, P.; Liu, Y.; Liu, J.Toxicity Mechanisms and Remediation Strategies for Chromium Exposure in the Environment. *Front. Environ. Sci.* **2023**, *11*, 1131204.

49  Sharma, P.; Singh, S. P.; Parakh, S. K.; Tong, Y. W. Health Hazards of Hexavalent Chromium (Cr (VI)) and its Microbial Reduction. *Bioengineered* **2022**, *13*(3), 4923-4938.

50  Xie, S. Water Contamination Due to Hexavalent Chromium and Its Health Impacts: Exploring Green Technology for Cr (VI) Remediation. *Green Chem. Lett. Rev.* **2024**, *17*(1), 2356614.

51  Roychoudhury, S.; Qiao, R.; Zhou, Z.; Li, Q.; Lyu, Y.; Kim, J.H.; Liu, J.; Lee, E.; Polzin, B.J.; Guo, J.; Yan, S. Deciphering the Oxygen Absorption Pre-edge: a Caveat on Its Application for Probing Oxygen Redox Reactions in Batteries. *Energy Environ. Mater.* **2021**, *4*(2), 246-254.

52  Wang, L.; Yang, Z.; Bowden, M.E.; Freeland, J.W.; Sushko, P.V.; Spurgeon, S.R.; Matthews, B.; Samarakoon, W.S.; Zhou, H.; Feng, Z.; Engelhard, M.H. Hole-Trapping-Induced Stabilization of $Ni^{4+}$ in $SrNiO_3/LaFeO_3$ Superlattices. *Adv. Mater.* **2020**, *32*(45), 2005003.

**Figures and Figure caption**

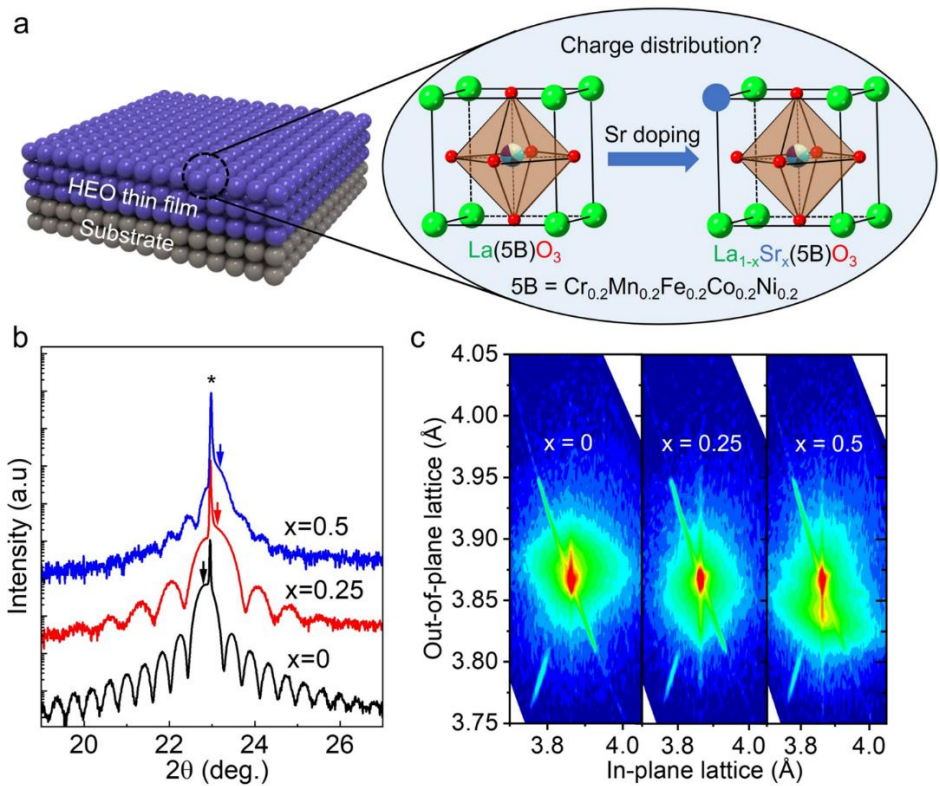

**Figure 1.** (a) Schematic illustration of the growth of HEPO $La_{1-x}Sr_x(5B)O_3$ thin films, highlighting the question of how Sr doping influences charge distribution. (b) High-resolution XRD θ–2θ scans for $La_{1-x}Sr_x(5B)O_3$ thin films grown on LSAT(001). The LSAT substrate peaks are marked with "*", while arrows denote the (001) diffraction peaks for $La_{1-x}Sr_x(5B)O_3$ films. (c) Reciprocal space maps near the (103) reflection of the substrate confirm coherent strain states.

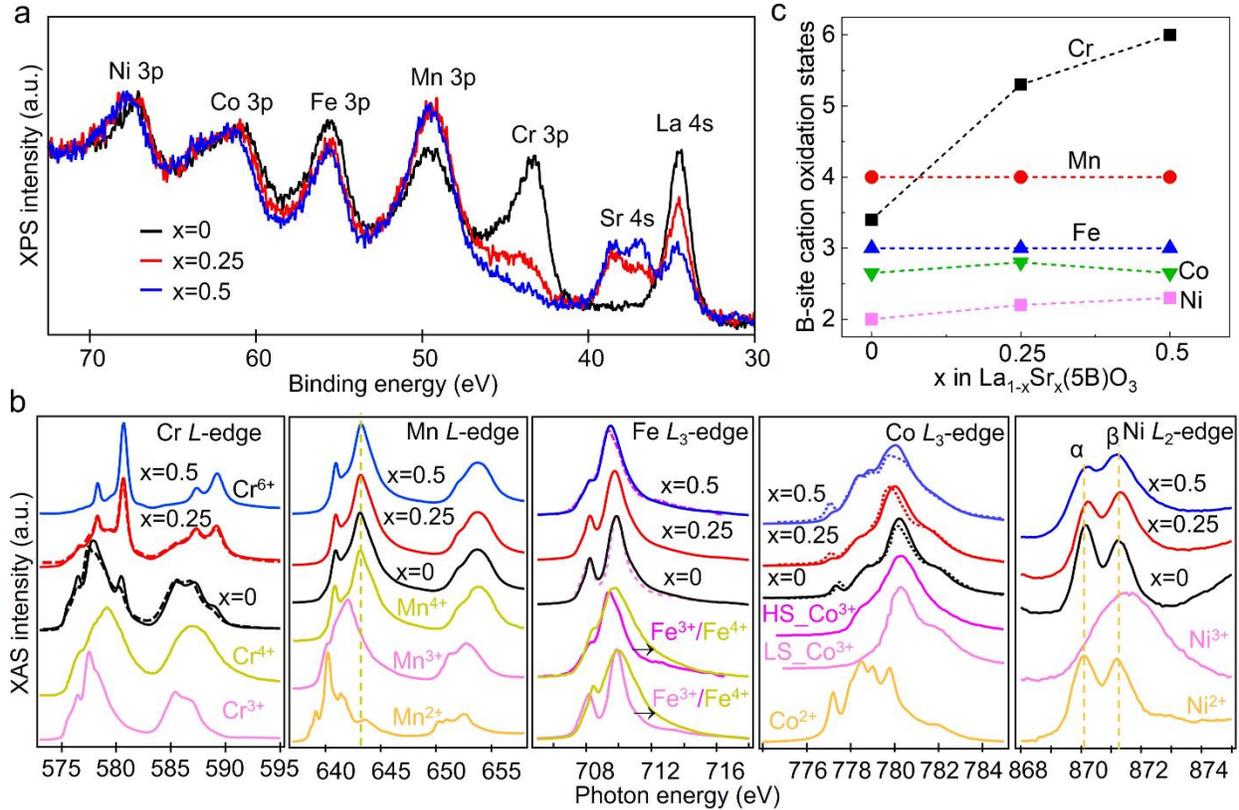

**Figure 2.** (a) *TM* 3p core-level spectra of HEPOs with varying Sr doping level (x = 0, 0.25, and 0.5). (b) XAS spectra at the Cr $L_{2,3}$ edges, Mn $L_{2,3}$ edges, Fe $L_3$ edge, Co $L_3$ edge, and Ni $L_3$ edge for HEPOs with different Sr doping levels. For Cr (Co), the black, red and blue dashed lines denote fitting results from a linear combination of $Cr^{3+}$, $Cr^{4+}$, and $Cr^{6+}$ ($Co^{2+}$, LS_$Co^{3+}$, and HS_$CO^{3+}$) reference spectra. $Cr^{3+}$ and $Cr^{4+}$ reference spectra were obtained from $Cr_2O_3$ and $CrO_2$, respectively.[51] Note that we selected the Cr spectra for x = 0.5 as the reference for $Cr^{6+}$, as it aligns well with the $Cr^{6+}$ spectra reported in the literature.[51] $Mn^{2+}$, $Mn^{3+}$, and $Mn^{4+}$ reference spectra were obtained from MnO, $LaMnO_3$, and $LaMn_{0.5}Co_{0.5}O_3$, respectively.[39] $Fe^{3+}$ and $Fe^{4+}$ reference spectra shown on the bottom (in the middle) were obtained from $LaFeO_3$ ($SrFeO_{2.5}$) and $(SrNiO_3)_1/(LaFeO_3)_1$ superlattice ($SrFeO_3$) for $Fe^{3+}$ and $Fe^{4+}$, respectively.[42,52] For comparison, we overlay the Fe $L_3$-edge spectrum of $LaFe^{3+}O_3$ ($SrFe^{3+}O_{2.5}$) on x = 0 (x = 0.5). $Co^{2+}$, LS_$Co^{3+}$, and HS_$CO^{3+}$ reference spectra were obtained from CoO, $EuCoO_3$, and $Sr_2CoO_3Cl$.[19] $Ni^{2+}$ and $Ni^{3+}$ reference spectra were obtained from NiO and $LaNiO_3$, respectively.[41] (d) Spectroscopically determined average B-site cation oxidation states as a function of x in $La_{1-x}Sr_x(5B)O_3$.

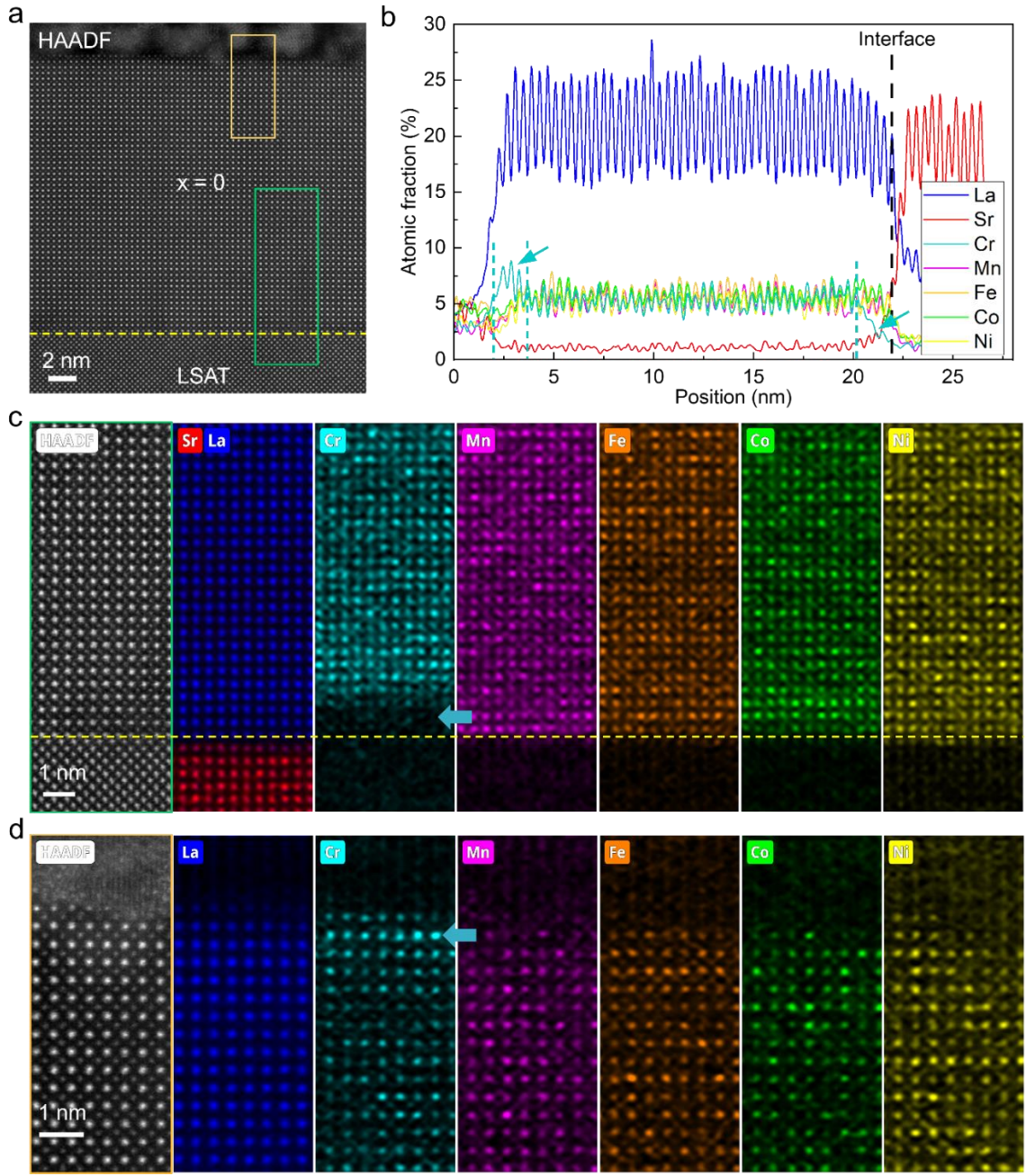

**Figure 3.** (a) HAADF-STEM image of the La(5B)O$_3$ (x = 0) film viewed along [100] direction of the LSAT substrate. (b) Integrated line profiles of atomic fractions, representing the spatial distribution of individual A-site and B-site cations in the ABO$_3$ structure. (c-d) Atomic-resolution EDS elemental maps showing the distribution of Sr/La, Cr, Mn, Fe, Co, and Ni at the film/substrate interface (c) and the film surface (d). The yellow dashed lines in (a) and (c) denote the film/substrate interface. Aqua arrows in (b-d) emphasize Cr depletion at the film/substrate interface and Cr enrichment at the film surface.

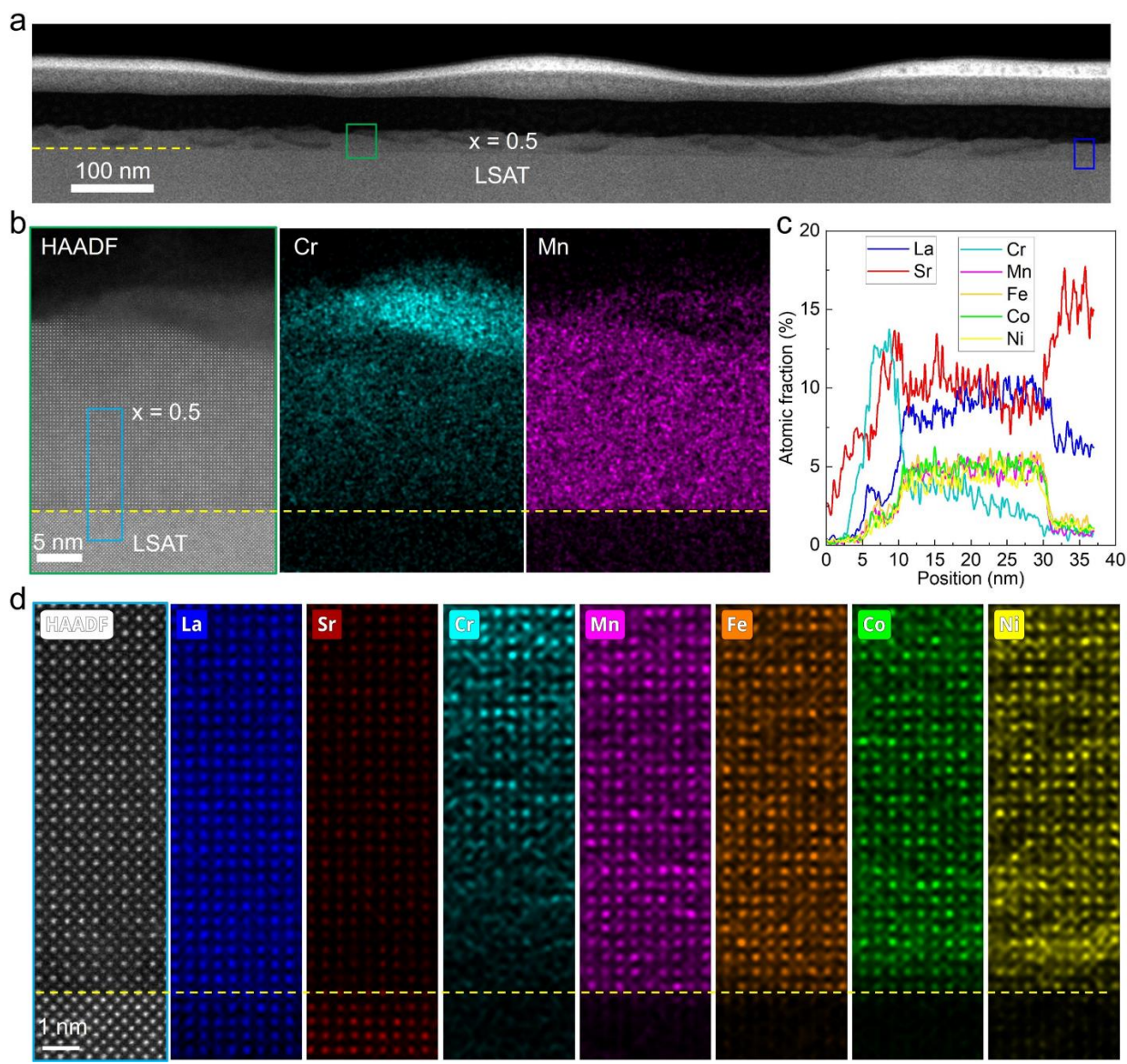

**Figure 4.** (a) Wide-field STEM image of $La_{1-x}Sr_x(5B)O_3$ with x = 0.5, viewed along [100] direction of the LSAT substrate. (b) Enlarged HAADF-STEM image of the region marked by the green box in (a), along with corresponding EDS elemental maps of Cr and Mn, highlighting Cr enrichment in the amorphous region on the film surface. (c) Integrated line profiles of atomic fractions, showing the spatial distribution of individual A-site and B-site cations in the $ABO_3$ structure. (d) Enlarged HAADF-STEM image of the region marked by the light blue box in (b), accompanied by atomic-resolution EDS elemental maps of La, Sr, Cr, Mn, Fe, Co, and Ni, revealing significant Cr depletion at the film/substrate interface.

TOC

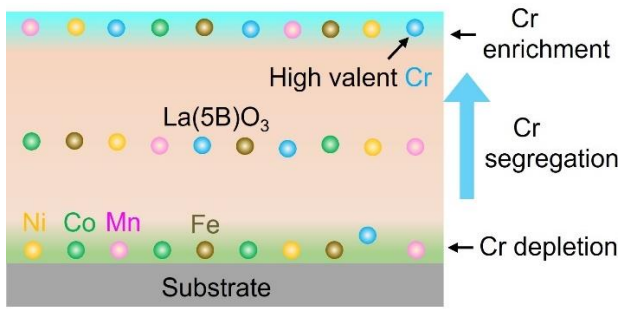

# Supporting information

# for

Le Wang[1,†,*], Krishna Prasad Koirala[1,†], Shuhang Wu[2], Jueli Shi[1], Hsin-Mei Kao[1,3], Andrew Ho[1,4], Minju Choi[1], Dongchen Qi[2], Anton Tadich[5], Mark E. Bowden[1], Bethany E. Matthews[6], Yang Yang[7], Hua Zhou[8], Chih-hung Chang[3], Zihua Zhu[9], Chongmin Wang[6], Yingge Du[1,*]

[1]Physical and Computational Sciences Directorate, Pacific Northwest National Laboratory, Richland, WA 99354, USA
[2]Centre for Materials Science, School of Chemistry and Physics, Queensland University of Technology, Brisbane, Queensland 4000, Australia
[3]School of Chemical, Biological and Environmental Engineering, Oregon State University, Corvallis, OR 97331, USA
[4]Santa Rosa Junior College, Santa Rosa, CA 95401, USA
[5]Australian Synchrotron, Clayton 3168, VIC, Australia
[6]Energy and Environment Directorate, Pacific Northwest National Laboratory, Richland, Washington, 99352 USA
[7]National Synchrotron Light Source II, Brookhaven National Laboratory, Upton, NY 11973, USA
[8]X-Ray Science Division, Advanced Photon Source, Argonne National Laboratory, Lemont, IL 60439, USA
[9]Environmental Molecular Sciences Laboratory, Pacific Northwest National Laboratory, Richland, WA 99354, USA

[†]These authors contributed equally to this work.

**Methods**

*Thin Film Growth*: Thin films of $La_{1-x}Sr_x(5B)O_3$ (where $5B = Cr_{0.2}Mn_{0.2}Fe_{0.2}Co_{0.2}Ni_{0.2}$) with x= 0, 0.25, and 0.5 were grown on (001)-oriented $(LaAlO_3)_{0.3}(Sr_2AlTaO_6)_{0.7}$ (LSAT) substrates by PLD using a KrF excimer laser ($\lambda$ = 248 nm) at a fluence of ~1.3 J cm$^{-2}$ and a repetition rate of 5 Hz. During the deposition, the substrate temperature was kept at 650 °C and the oxygen partial pressure was maintained at 150 mTorr. After deposition, all the $La_{1-x}Sr_x(5B)O_3$ films were cooled down to room temperature in 150 mTorr of $O_2$.

*Sample Characterization*: The lattice structure was characterized by a high-resolution XRD (Rigaku SmartLab) using a rotating anode Cu K$\alpha$ source. TEM samples were prepared using a Helios FIB-SEM system following a standard lift-out process. A cross-sectional lamella was extracted via FIB milling, initially thinned to about 200 nm at 30 kV. Subsequent milling steps at 5 kV and 2 kV further reduced the thickness to approximately 50 nm, with final polishing performed at 2 kV. STEM-EDS was performed with a Thermo Fisher Scientific Spectra Ultra system featuring an aberration corrector and Ultra-X EDS detector, operating at 300 kV. To refine

EDS data and resolve overlapping peaks, the Filtered Least Squares method in Velox software was used. Both STEM-EDS mapping and EELS were carried out at 300 kV with a probe current of approximately 40 pA. Dual-EELS mode was employed for energy calibration, simultaneously capturing the zero-loss spectrum alongside core-loss spectra. To improve signal-to-noise ratio, point EELS spectra were collected from individual Cr/Fe atomic columns, with 20 frames summed. HAADF-STEM imaging was performed with a 30 mrad convergence angle and a collection angle of 68–180 mrad.

*XPS, XAS, and ToF-SIMS Measurements*: To avoid the sample charging issue, high-resolution XPS using a monochromatic Al-Kα X-ray source was carried out on all as-grown samples. Cr *L*-edge, Mn *L*-edge, Fe *L*-edge, Co *L*-edge, and Ni *L*-edge XAS were measured at the Soft X-ray Spectroscopy beamline at the Australian Synchrotron. Fe *K*-edge, Co *K*-edge, and Ni *K*-edge XAS were measured at 5-ID SRX (Sub-micron Resolution X-ray Spectroscopy) beamline at NSLS-II. Time-of-flight secondary ion mass spectrometry (ToF-SIMS) measurements were taken using a ToF-SIMS V mass spectrometer (ION-TOF, GmbH, Münster, Germany). A dual beam, interlaced mode strategy was used, wherein positive spectra data collection was taken using a 25 keV Bi+ analysis beam (~1.14 pA, 100 μm × 100 μm scanning area) while depth profiling was enabled via a 1 keV $O_2^+$ sputtering beam (~239 nA, 300 μm × 300 μm scanning area). Additionally, a low energy flood gun (10 eV, ~ 2.58 μA) was used for charge compensation. The film/substrate interface was determined via the secondary ion signals of $Fe^+$ and $Al^+$. Prior to each measurement, the analysis beam was centered in the middle of the sputtering crater to avoid any edge-related effects. All XPS, XAS, and ToF-SIMS measurements were conducted at room temperature.

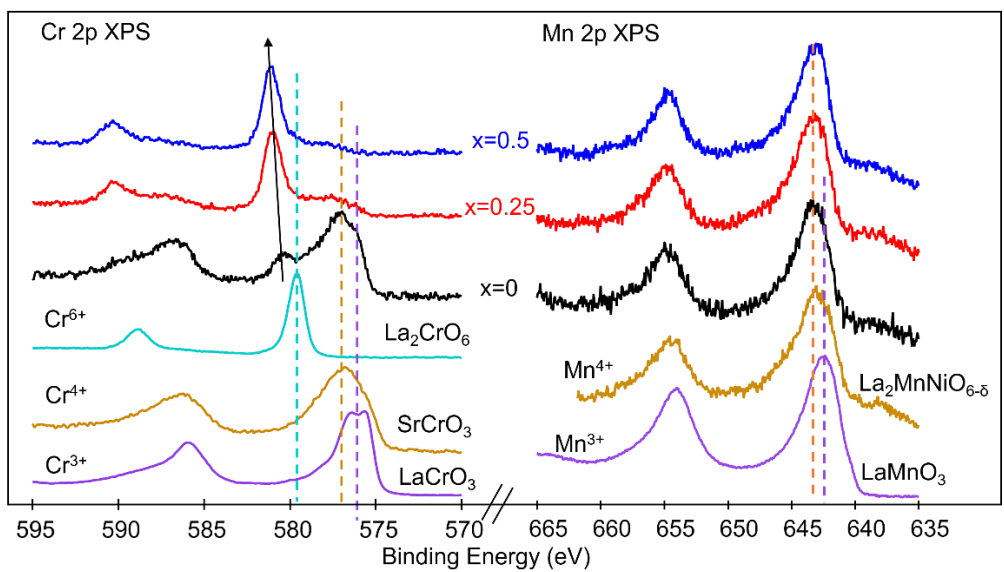

**Figure S1.** Cr 2p and Mn 2p XPS. The dashed purple and yellow lines denote the Cr (Mn) 2p binding energy for LaCr$^{3+}$O$_3$ (LaMn$^{3+}$O$_3$) and SrCr$^{4+}$O$_3$ (La$_2$Mn$^{4+}$NiO$_{6-\delta}$), respectively.[S1, S2] The aqua dashed line denotes the Cr 2p binding energy for La$_2$Cr$^{6+}$O$_6$.[S1] For the x = 0 sample, the Cr 2p XPS analysis suggests that Cr exists as a mixed oxidation state comprising Cr$^{3+}$, Cr$^{4+}$, and Cr$^{6+}$. With Sr doping, the proportion of highly oxidized Cr$^{6+}$ increases, while the amounts of Cr$^{3+}$ and Cr$^{4+}$ decrease. For Mn, all La$_{1-x}$Sr$_x$(5B)O$_3$ thin films exhibit identical Mn 2p spectra, indicating that the Mn oxidation state remains unchanged with Sr doping.

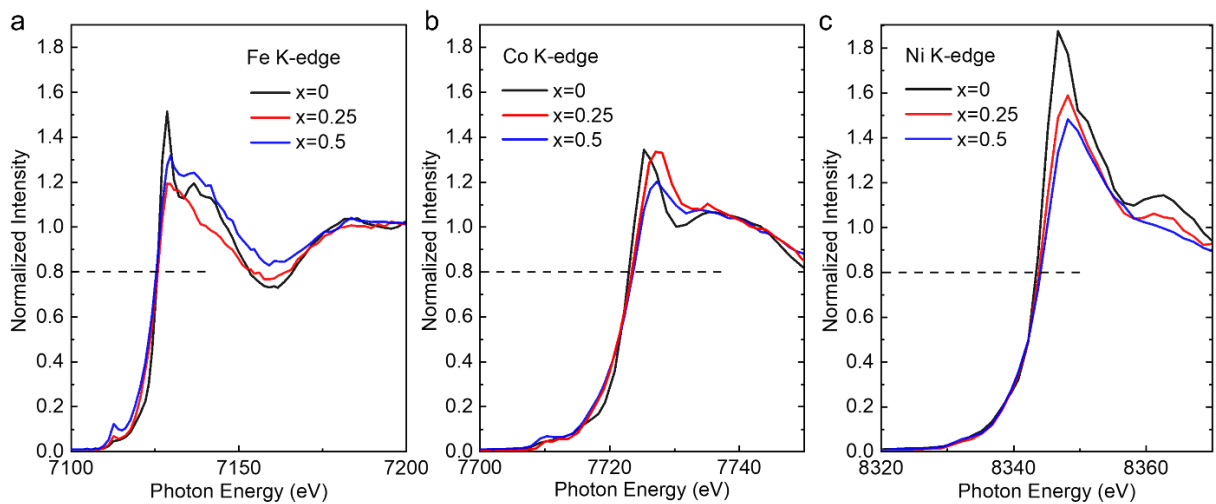

**Figure S2.** (a) Fe *K*-edge, (b) Co *K*-edge, (c) Ni *K*-edge XAS. According to previous studies,[S3-S5] the energy corresponding to normalized absorption of ≈0.8 (indicated by black dashed lines) can be used to determine the *TM* oxidation state. The Fe K-edge spectra for all three samples are located at the same energy position, indicating that the Fe oxidation state remains unchanged. In contrast, the energy positions of Co K-edge and Ni K-edge spectra shift toward higher photon energy with increasing Sr doping levels, suggesting an increase in the oxidation states of Co and Ni with Sr doping.

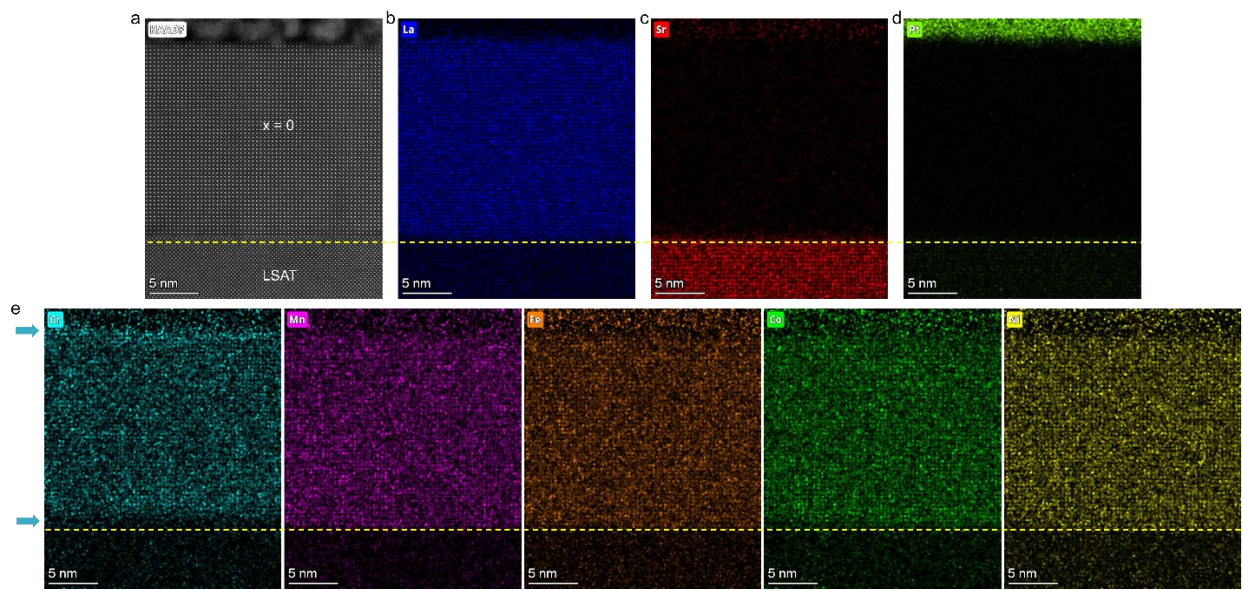

**Figure S3.** (a) High resolution HAADF-STEM image of the La(5B)O$_3$/LSAT interface. (b-e) EDS elemental maps acquired from the same region as the HAADF-STEM image in panel (a). The yellow dashed lines indicate the interface, while the aqua arrows highlight the Cr depletion at the film/substrate interface and the Cr enrichment at the film surface.

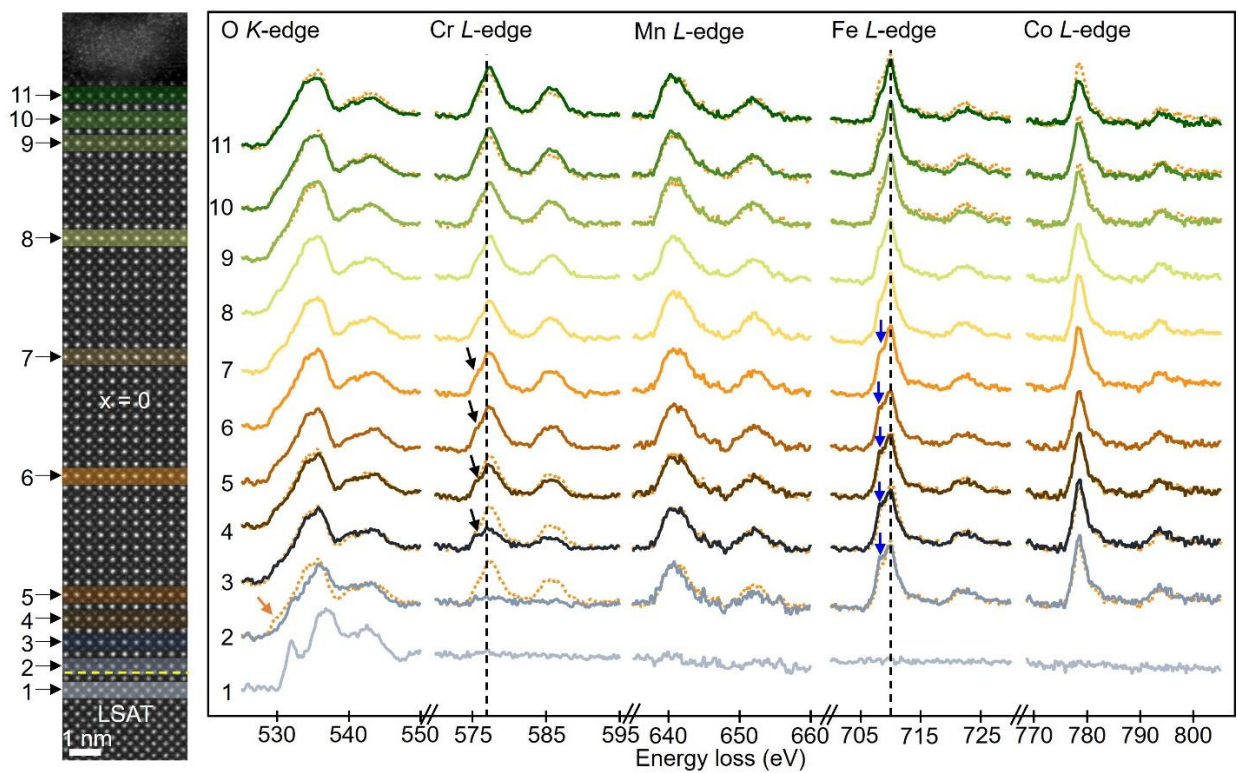

**Figure S4.** Layer-resolved EELS analyses show a gradual increase in the Cr oxidation state from $Cr^{3+}$ at the interface to a higher state (remaining below $Cr^{4+}$, as indicated by peak shifts of less than 0.5 eV) toward the surface. Among other cations, only Fe exhibits notable variations in its L-edge spectral shape, with a low-energy shoulder (marked by blue arrows) near the interface that may indicate the presence of $Fe^{2+}$. Nonetheless, soft and hard XAS confirm that Fe remains $Fe^{3+}$, suggesting that changes in structural coordination—rather than oxidation state—account for the observed Fe L-edge differences. This conclusion is further supported by O K-edge EELS, where only region 1 (the Cr-depleted region) exhibits distinct features compared to other regions. Additionally, Co appears to play a stabilizing role during Cr migration, as demonstrated by its higher concentration near the interface and lower concentration at the surface, as highlighted in the TOC figure.

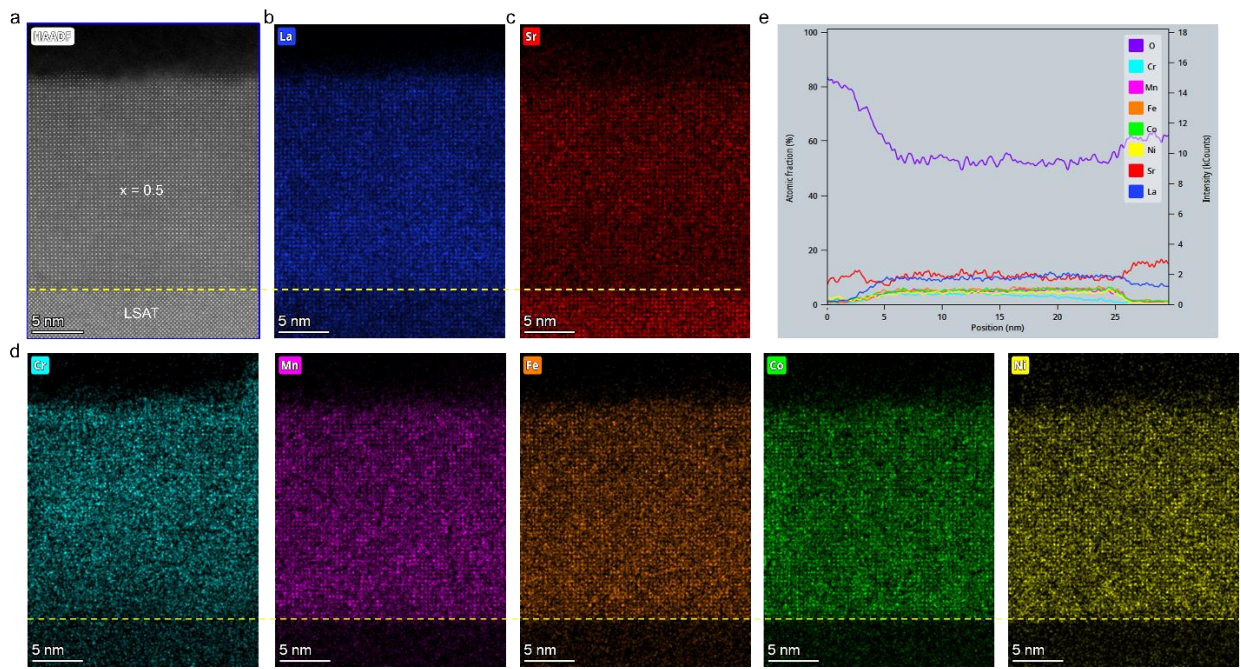

**Figure S5.** (a) High resolution HAADF-STEM image of the La$_{0.5}$Sr$_{0.5}$(5B)O$_3$/LSAT interface. (b-d) EDS elemental maps acquired from the same region as the HAADF-STEM image in panel (a). The yellow dashed lines indicate the interface. (e) Integrated line profiles of atomic fractions.

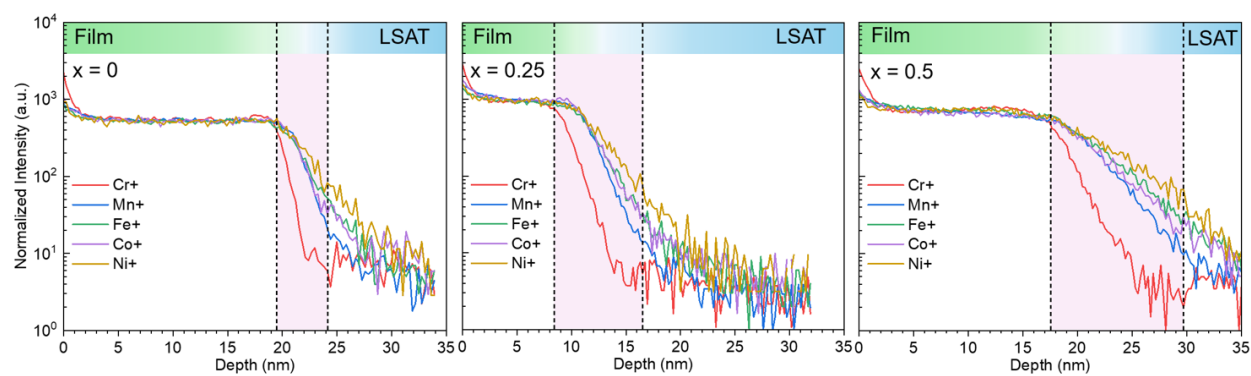

**Figure S6.** ToF-SIMS depth profiles for all three $La_{1-x}Sr_x(5B)O_3$ samples reveal surface Cr enrichment and interfacial Cr depletion. The pink-highlighted regions indicate the thickness of the Cr depletion layer across samples. This layer is defined as the region where the Cr concentration drops to 20% below that of the other four B-site elements and extends until the depth which the average concentration of the other four species exceeds five times the Cr baseline.

**References**

[S1] Qiao. L.; Xiao. H. Y.; Heald, S. M.; Bowden, M. E.; Varga. T.; Exarhos. G. J.; Biegalski, M. D.; Ivanov. I. N.; Weber, W. J.; Droubay, T. C.; Chambers, S. A. The impact of crystal symmetry on the electronic structure and functional properties of complex lanthanum chromium oxides. *J. Mater. Chem. C* **2013**, 1, 4527-4535.

[S2] Spurgeon, S.R.; Du, Y.; Droubay, T. C.; Devaraj, A.; Sang, X.; Longo, P.; Yan, P.; Kotula, P. G.; Shutthanandan, V.; Bowden, M. E.; LeBeau, J. M.; Wang, C.; Sushko, P. V.; Chambers, S. A. Competing Pathways for Nucleation of the Double Perovskite Structure in the Epitaxial Synthesis of $La_2MnNiO_6$. *Chem. Mater.* **2016**, 28(11), 3814-3822.

[S3] Adiga, P.; Wang, L.; Wong, C.; Matthews, B.E.; Bowden, M.E.; Spurgeon, S.R.; Sterbinsky, G.E. Blum, M.; Choi, M.J.; Tao, J.; Kasper, T.C.; Chambers, S.A.; Stoerzinger, K.A.; Du, Y. Correlation Between Oxygen Evolution Reaction Activity and Surface Compositional Evolution in Epitaxial $La_{0.5}Sr_{0.5}Ni_{1-x}Fe_xO_{3-\delta}$ Thin Films. *Nanoscale* **2023**, *15*(3) 1119-1127.

[S4] Zhang, H.; Gao, Y.; Xu, H. ; Guan, D.; Hu, Z.; Jing, C.; Sha, Y.; Gu, Y.; Huang, Y.C.; Chang, Y.C.; Pao, C.W. Combined Corner-Sharing and Edge-Sharing Networks in Hybrid Nanocomposite with Unusual Lattice-Oxygen Activation for Efficient Water Oxidation. *Adv. Funct. Mater.* **2022**, *32*(45), 2207618.

[S5] Guan, D.; Zhang, K.; Hu, Z.; Wu, X.; Chen, J.L.; Pao, C.W.; Guo, Y.; Zhou, W.; Shao, Z. Exceptionally Robust Face-Sharing Motifs Enable Efficient and Durable Water Oxidation. *Adv. Mater.* **2021**, *33*(41), 2103392.